\documentclass[aps,prb,twocolumn]{revtex4}

\usepackage{graphicx,color}
\usepackage{amssymb,amsfonts,amsmath}

\def\dir{.}
\def\co{c_{_0}}
\def\dc{\delta c_{_0}}

\newcommand{\lo}{{\em lo}}
\newcommand{\ld}{{\em ld}}
\newcommand{\PP}{{\em P }}
\newcommand{\CC}{{\em C }}

\newcommand{\MPP}{{\em \footnotesize P }}
\newcommand{\MCC}{{\em \footnotesize C }}

\newcommand{\degree}{\mbox{${}^0$}}

\begin{document}

\title{Monolayer curvature stabilizes nanoscale raft domains in mixed lipid bilayers}

\author{Sebastian Meinhardt${}^1$, Richard L.C. Vink${}^2$, 
        Friederike Schmid${}^1$}
\affiliation{
${}^1$ Johannes Gutenberg-Universit\"at Mainz, Germany \\
${}^2$ Georg-August-Universit\"at G\"ottingen}

\begin{abstract} According to the lipid raft hypothesis, biological lipid 
membranes are laterally heterogeneous and filled with nanoscale ordered ``raft'' 
domains, which are believed to play an important role for the organization of 
proteins in membranes. However, the mechanisms stabilizing such small rafts are 
not clear, and even their existence is sometimes questioned. Here we report the 
observation of raft-like structures in a coarse-grained molecular model for 
multicomponent lipid bilayers. On small scales, our membranes demix into a 
liquid ordered (\lo) and a liquid disordered (\ld) phase. On large scales, phase 
separation is suppressed and gives way to a microemulsion-type state that 
contains nanometer size \lo~domains in a \ld~environment. Furthermore, we 
introduce a mechanism that generates rafts of finite size by a coupling between 
monolayer curvature and local composition. We show that mismatch between the 
spontaneous curvatures of monolayers in the \lo~and \ld~phase induces elastic 
interactions, which reduce the line tension between the \lo~and \ld~phase and 
can stabilize raft domains with a characteristic size of the order of a few 
nanometers. Our findings suggest that rafts in multicomponent bilayers might be 
closely related to the modulated ripple phase in one-component bilayers. 
\end{abstract}



\maketitle

Ever since its introduction some two decades ago 
\cite{SimonsvanMeer_88, SimonsIkonen_97}, the lipid raft concept has been 
discussed controversially \cite{ZurzolovanMeer_03, Pike_06, 
Hancock_09,Leslie_11}.  It rests on two established facts: (i) Biological 
membranes are laterally heterogeneous. Heterogeneity is necessary to achieve the 
functions of membrane proteins, e.g., in cellular signal transduction and 
trafficking or in endocytosis \cite{VerebSzollosi_03}.  (ii) Lipid-lipid phase 
separation is observed in model multicomponent lipid bilayers. A variety of 
ternary mixtures containing cholesterol phase separate into a cholesterol-poor 
$L_\alpha$ or ``liquid disordered'' (\ld) phase and a cholesterol-rich ``liquid 
ordered" (\lo) phase with a higher degree of chain order \cite{VeatchKeller_05}. 
The lipid raft hypothesis states that lipid-lipid phase separation contributes 
to membrane heterogeneity and is exploited by nature to organize proteins 
\cite{BrownLondon_98, JacobsonDietrich_99, Edidin_03, Pike_03, SimonsVaz_04, 
ZeydaStulnig_06, Hanzal-BayerHancock_07, LingwoodSimons_10}.  Preexisting 
``raft domains'' supposedly provide a heterogeneous environment that sorts 
proteins and brings them close to each other, thus facilitating the 
protein-protein interaction needed for clustering. This concept provides an 
elegant picture for a number of experimental observations, e.g., the reduced 
mobility of raft-associating proteins, which depends on cholesterol content 
\cite{PralleKeller_00}, the submicroscopic local clustering of raft-associated 
proteins as observed by fluorescence resonance energy transfer 
\cite{ZachariasViolin_02}, or the existence of detergent resistant membrane 
fragments (DRM) with a high content of cholesterol, sphingolipids, and 
raft-associated proteins \cite{BrownRose_92}.

However, all the experimental evidence is rather indirect and the
interpretation in terms of lipid rafts has been subject to debate
\cite{Leslie_11}.  One problem with lipid rafts is that they cannot be observed
in vivo with optical microscopic techniques, hence they must be tiny. Diffusion
experiments indicate that the domain sizes of rafts are probably in the range
of a few tens of nanometers \cite{PralleKeller_00}, and they have short
lifetimes in the millisecond or microsecond regime.  Thus the current view is
that rafts constitute dynamically changing nanoscale entities, which are
collected to larger arrays upon need by protein-protein interactions
\cite{Hancock_09, Edidin_03, LingwoodSimons_10}.

This naturally raises questions regarding the physical nature of rafts and the 
mechanisms stabilizing them. The first question is: If the physical basis of 
rafts is phase separation, why are they so small? A number of possible 
explanations have been pointed out. For example, it was argued that membranes 
{\em in vivo} are not at thermodynamic equilibrium, and the constant turnover of 
lipids may well disrupt the formation of large phase separated domains 
\cite{TurnerSens_05}. Alternatively, it was proposed that immobilized cytoplasm 
proteins generate disorder in the membrane which prevents large scale phase 
separation \cite{YethirajWeisshaar_07}.

The second important question is whether biological membranes really do tend to 
phase separate. Model multicomponent membranes may exhibit phase separation at 
physiological temperatures \cite{VeatchKeller_03, VeatchKeller_05b}. However, it 
is not clear whether this is also true for biological membranes. Veatch et al.\ 
\cite{VeatchCicuta_08} have isolated giant plasma vesicles directly from living 
rat cells and showed that they do undergo a demixing phase transition, but the 
demixing temperature is around $T \sim 15-25\degree$C. Hence such membranes 
would be in a mixed state at physiological temperatures. Veatch et al.\cite{VeatchSoubias_07} and Honerkamp-Smith et al.\cite{Honerkamp-SmithVeatch_09} have argued that raft-like structures might emerge as a signature of 
critical fluctuations. This 
would restrict ``rafts'' to relatively small regions in parameter space, since 
critical clusters only become large close to critical points.

Alternatively, the membrane might be in the state of a two dimensional 
microemulsion, which is globally homogeneous, but locally phase separated with a 
characteristic length scale or domain size \cite{GompperSchick_book}. Schick 
\cite{Schick_12} has recently proposed a mechanism that would stabilize a 
microemulsion, which builds on a coupling between the local curvature of the 
bilayer and the local composition difference between the two 
leaflets\cite{SafranPincus_90}. Such a coupling can generate modulated phases in 
mixed membranes under tension \cite{LeiblerAndelman_87, HardenMacKintosh_94, 
KumarGompper_99, BaumgartHess_03}. Schick argued that it could also stabilize a 
microemulsion-type state with characteristic length scale of order 100~nm in 
membranes under tension, which however would diverge in tensionless membranes.

A more traditional idea is that the membranes contain line-active agents 
\cite{SimonsVaz_04} which reduce the line tension and eventually turn a 
phase-separated mixture into a microemulsion. 
\begin{figure}
\centerline{\includegraphics[width=.2\textwidth]{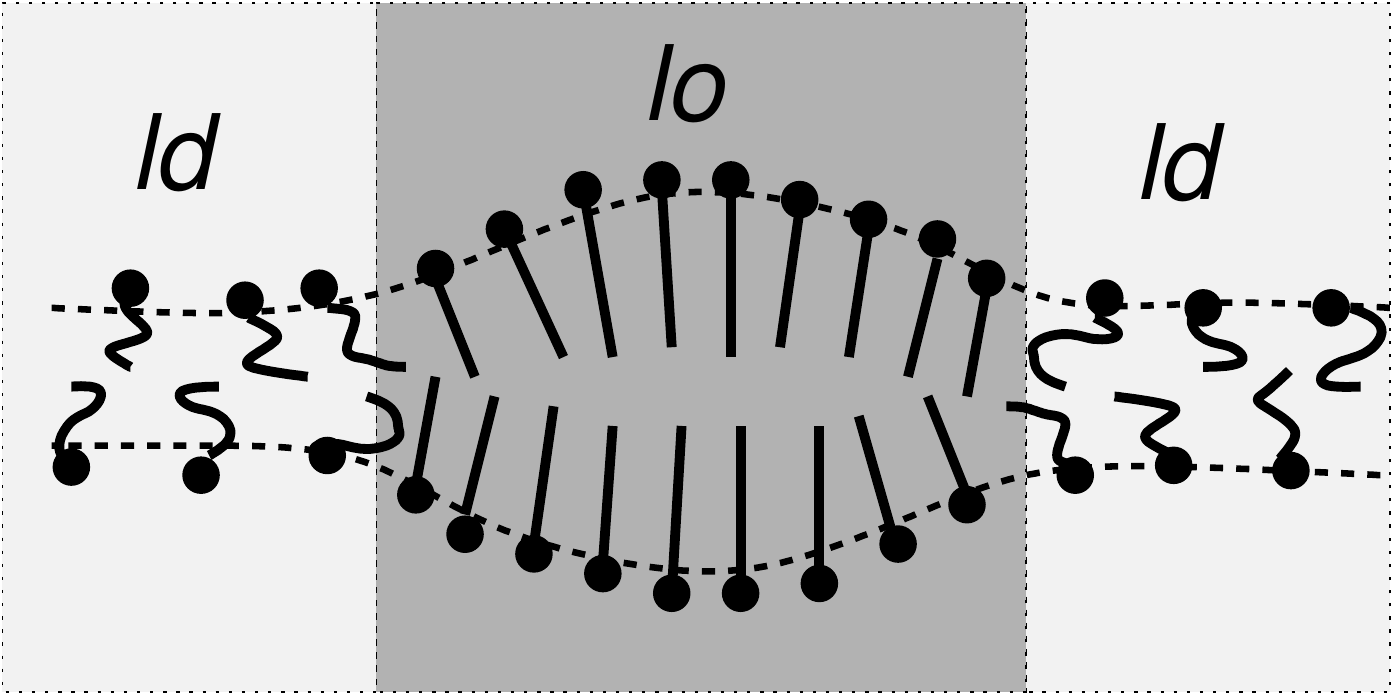}}
\caption{Curvature mechanism generating rafts. The liquid ordered (\lo) and 
liquid disordered (\ld) regions in opposing monolayer leaflets are spatially 
correlated and have different spontaneous curvatures. This stabilizes domains of 
a finite size.} \label{fig:raft_cartoon}
\end{figure}
\noindent 
Possible candidate agents are proteins \cite{Hancock_09} or minority lipids 
with a saturated and an 
unsaturated tail \cite{BrewsterPincus_09, YamamotoBrewster_10}. However, none of 
the lipids that are considered typical for lipid rafts, i.e., cholesterol or 
sphingolipids, have obvious line-active properties \cite{Schick_12}. Thus this 
mechanism relies on additional assumptions regarding the composition of 
membranes containing rafts.

Typical model membranes for studying rafts (``raft mixtures'') contain 
cholesterol and at least two other lipid components. Three components seem 
necessary to bring about global lateral phase separation between liquid membrane 
phases~\cite{VeatchKeller_05}. However, there is some evidence that nanoscopic 
domains may already be present in binary mixtures containing cholesterol, in 
particular, in mixtures of cholesterol and a lipid with high main transition temperature ($T_m$). The 
literature on these mixtures is controversial. Whereas several authors have 
observed immiscible liquid phases, based on various techniques such as ESR, NMR, 
or diffusivity experiments \cite{IpsenKarlstrom_87, SankaramThompson_91, 
AlmeidaVaz_92, LindblomOradd_09}, others claim that cholesterol and lipids are 
miscible in the whole high temperature fluid range (see Ref.\ 
\cite{VeatchKeller_05} and references therein). In particular, fluorescence 
microscopy images feature only one homogeneous phase~\cite{VeatchKeller_05}.  
Feigenson \cite{Feigenson_09} has introduced the notion of ``type~I'' and 
``type~II'' mixtures, where the ``type~II mixtures'' exhibit global phase 
separation, whereas ``type~I mixtures'' phase separate on the nanoscale, but are 
globally homogeneous, much like microemulsions. The experimental evidence 
suggests that binary lipid-cholesterol mixtures might be type~I mixtures. 
Thus bilayers of binary lipid-cholesterol mixtures already seem to have many of 
the intriguing properties attributed to lipid rafts, and a theoretical study of 
such binary systems should provide insight into the mechanisms stabilizing 
rafts.

In the present paper, we contribute to the raft discussion with two main 
results: First, we present Monte Carlo simulations of a coarse-grained molecular 
model for binary lipid bilayers which demonstrate the existence of a 
thermodynamically stable heterogeneous membrane phase with raft-like 
\lo~nanodomains in a \ld~environment. Hence raft formation is found to be a 
generic phenomenon which can already be observed in binary mixtures and does not 
require specific line-active agents. Second, we present a theory which 
rationalizes our results and explains raft formation by a coupling between local 
composition and monolayer curvature. The theoretical picture is 
illustrated in Fig.~1. Liquid ordered domains have a 
propensity to bend inwards. If they oppose each other -- as suggested by 
simulations and also by experiments \cite{CollinsKeller_08} -- the bending 
competes with bilayer compression. This creates elastic tension which reduces 
the line tension even for tensionless membranes, and stabilizes domains with a 
well-defined diameter of the order of 10 nanometers.

Thus we identify a generic mechanism of raft formation in multicomponent 
membranes. It is different from the curvature-mediated mechanism proposed by 
Schick \cite{Schick_12}, which is based on bilayer curvature. Both Schick's and 
our mechanism rely on a competition between bending and an opposing force. If 
the bilayer as a whole has a propensity to bend, the opposing force is the 
surface tension. Therefore, the surface tension sets the characteristic length 
scale which thus diverges in tensionless membranes. In our monolayer curvature 
mechanism, the bending is opposed by bilayer compression, which results in a 
characteristic length scale of the order of the membrane thickness.
\section{Coarse-grained simulations of mixed lipid bilayers}
Our generic coarse-grained simulation model is based on a successful model for 
one-component lipid bilayers \cite{SchmidDuechs_07}, which reproduces the main 
phases of phospholipid membranes (liquid, tilted gel, and ripple phase) 
\cite{LenzSchmid_07, WestSchmid_10}, and the elastic properties of fluid 
dipalmitoylphosphatidylcholine (DPPC) bilayers at a semi-quantitative level 
\cite{WestBrown_09}. Here we introduce two types of lipids, ``phospholipids'' 
\PP and ``cholesterol'' \CC, with interactions designed such that \CC is smaller 
and stiffer than \PP and has a special affinity to \PP, reflecting the 
experimental observation that sterols in lipid bilayers always tend to be 
solubilized with one or two other lipids \cite{LindblomOradd_09}. For the 
purpose of the present study, it is essential to design the model such that it 
captures those non-random mixing effects, since they most likely drive the local 
segregation into a \ld~and a \lo~phase. Waheed \cite{Waheed_12} and Waheed et al. \cite{WaheedTjornhammar_12} recently studied the 
chemical potential of cholesterol in small systems of DPPC/cholesterol bilayers 
by atomistic (united-atom) and coarse-grained simulations. In 
atomistic simulations, they found that the chemical potential drops with the 
cholesterol concentration, indicating a clear tendency of local segregation into 
a cholesterol rich and a cholesterol poor phase [about 0.3 thermal enrgy units ($k_B T$) per lipid 
molecule]. This property was not reproduced by coarse-grained models that show 
no sign of random mixing \cite{WaheedTjornhammar_12}. Our model is described in 
more detail in the Methods section.

To examine whether the system phase separates locally, we first consider small 
systems (162 lipids) at fixed composition. We find that such small systems 
almost always assume one of two states, either disordered (\ld) or ordered 
(\lo), or jump between the two, depending on the composition. 
Figs.~2~(b) and (c) show sample configurations of these two 
states. For future reference we have evaluated the pressure profiles across 
monolayers in the two states and computed the spontaneous curvature of 
monolayers $c_0$ from the first moment \cite{Safran_book}. In the \ld~state we 
obtain $c_0=0.2 \pm 0.2 \sigma^{-1}$, and in the \lo~state, $c_0=1.22 \pm 0.09 
\sigma^{-1}$, using our simulation length unit $\sigma \sim 0.6$~nm. Thus 
monolayer regions in the \lo~state have a strong tendency to bend inwards, 
whereas monolayer regions in the \ld~state tend to remain flat.

The free energy gain $\mu$ for replacing a \PP chain by a \CC chain (the 
chemical potential difference) is shown in Fig.~2~(a). For 
temperatures above the main transition of the pure \mbox{\PP-system} ($T_m = 1.2 
\epsilon$), the chemical potential curve has an upward slope in the region 
between $\approx 10\%$ and $\approx 25\%$ \CC chains. This indicates an unstable 
regime where one would expect spontaneous demixing in larger systems.

When looking at larger systems (20.000 lipids), however, we find that the system 
does {\em not} phase separate globally. Instead, finite \lo~domains with a high 
concentration of \CC lipids appear, surrounded by the \ld~phase almost devoid of 
\CC lipids. To ensure that these domains are true equilibrium structures and not 
the result of incomplete phase separation, the simulations were conducted in the 
semi-grandcanonical ensemble at fixed $\mu$, i.e., lipids were allowed to switch 
their identities during the simulation. Figs.~3~(a) and (b) 
show an example of an equilibrated configuration. One can see the raft-like 
structure of the membrane from the top view (a) and the structure of alternating 
\lo~and \ld~regions from the side view (b). Consistent with this observation, 
the behavior of the \CC concentration as a function of $\mu$ shows no sign of a 
phase transition [Fig.~3~(c)].

\begin{figure}
\centerline{\includegraphics[width=.4\textwidth]{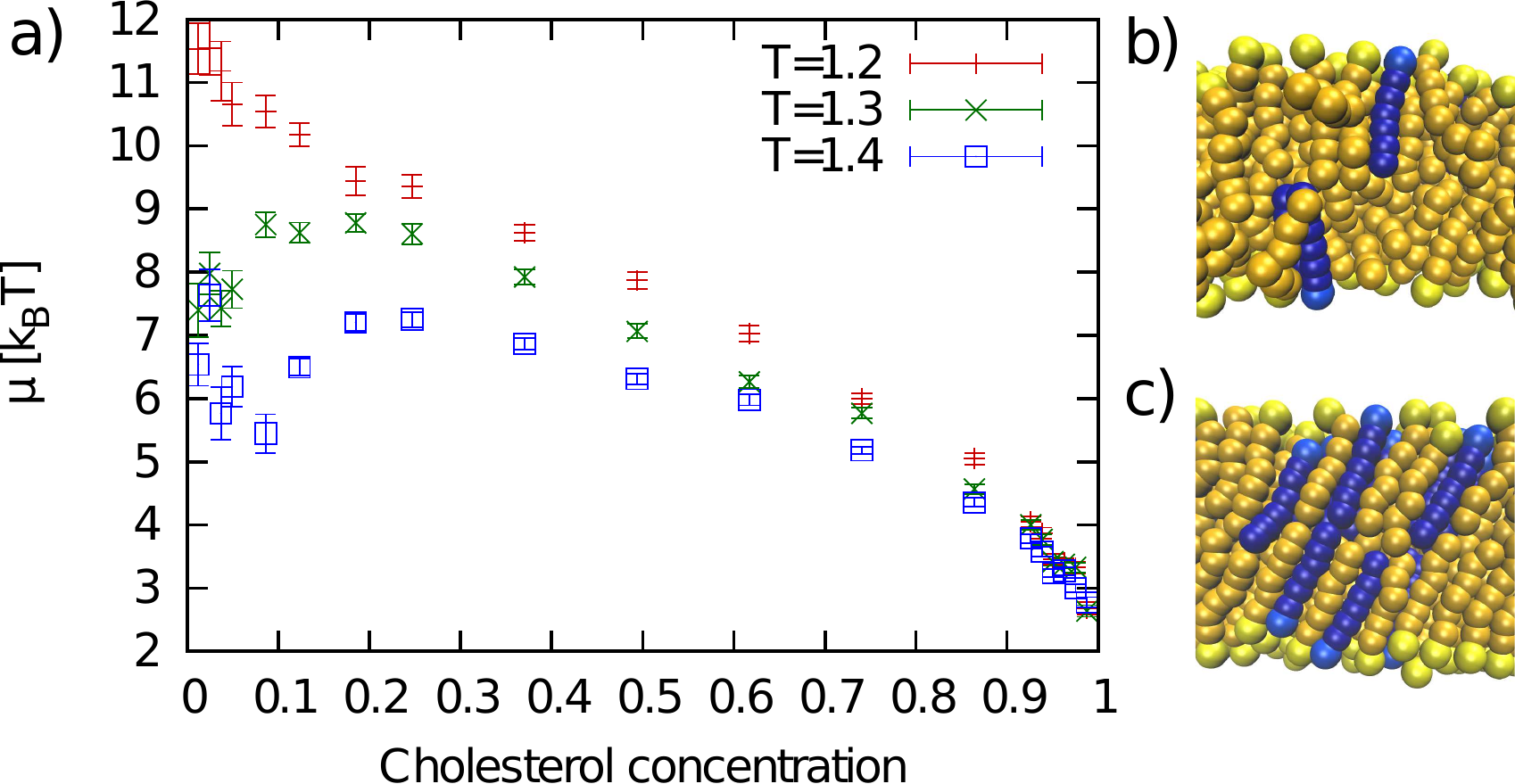}}
\caption{(a) Chemical potential difference in thermal energy units ($k_BT$) vs. cholesterol content 
from canonical simulations of small mixed bilayer systems (162 lipids). The 
corresponding snapshots show the system in the \ld~state (b) and the \lo~state 
(c). The darker chains represent \CC lipids.} \label{fig:small}
\end{figure}

\begin{figure}
\centerline{\includegraphics[width=.4\textwidth]{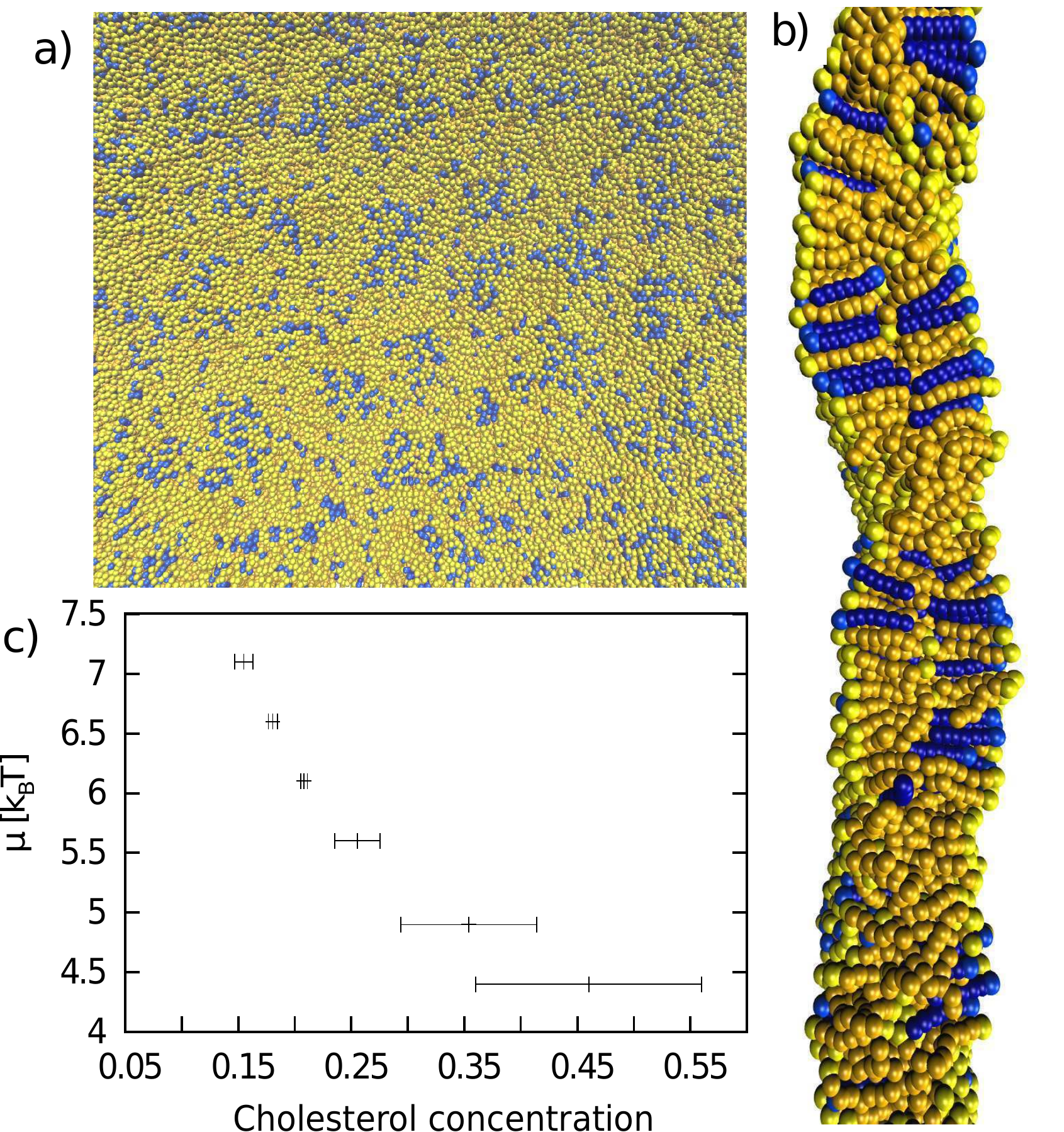}}
\caption{Snapshots of a large bilayer system (20.000 lipids) featuring raft-like 
\lo~domains: (a) top view, and (b) an enlarged section of a side view. 
Parameters are {\footnotesize $k_B T=1.4$, $\mu\approx 6.6 \, k_bT$}. (c) The 
chemical potential difference versus \CC (``cholesterol'') content from 
semi-grandcanonical simulations of large systems.}
\label{fig:snap_large}
\end{figure}

We analyze configurations such as shown in Fig.~3 following 
an algorithm described in the Methods section. This allows us to calculate the 
distribution of raft sizes (radii of gyration), shown in 
Fig.~4~(a). While the distributions look rather similar 
for different values of $\mu$, the area fraction (inset) clearly shows that in 
systems with larger $\mu$, i.e., systems with a lower concentration of \CC, a 
greater fraction of the total raft area is present in the form of smaller rafts.

Next, we address the question whether rafts in opposing monolayers are 
correlated. The normalized cross-correlation $K_i$ (see Methods section) between 
monolayers of a configuration $i$ tends to be positive with values ranging from 
$K=-0.05$ to $K=0.15$. To analyze whether a correlation $K_i > 0$ or 
anticorrelation $K_i < 0$ is significant, we generate a set of configurations 
with randomly shifted monolayers and determine the fraction of them which have a 
higher (anti)correlation $|\tilde{K}_i|$. At large $\mu$, i.e., small \CC 
concentrations, the distribution of $K_i$ is symmetric around zero and shifting 
monolayers often enhances the correlation. Raft domains are thus uncorrelated in 
this regime. At large \CC concentrations, however, all values of $K_i$ are 
positive and shifting monolayers almost always reduces the correlation. We 
conclude that the cross-correlation is significant at higher \CC concentrations, 
and that rafts tend to oppose each other in the bilayer. This is compatible with 
experimental observations in membranes exhibiting global phase separation, where 
it was found that \lo~domains are strongly correlated across the membrane 
\cite{CollinsKeller_08}.

Finally, we consider the in-plane structure factor 
[Fig.~5]. For small values of $\mu$ we observe a peak 
at nonzero $q \sim 0.05 \sigma^{-1}$, corresponding to a characteristic length 
scale of about $20 \sigma \sim 12$~nm. Such a peak is a typical signature of a 
microemulsion~\cite{GompperSchick_book}.

\begin{figure}
\centerline{\includegraphics[width=.4\textwidth]{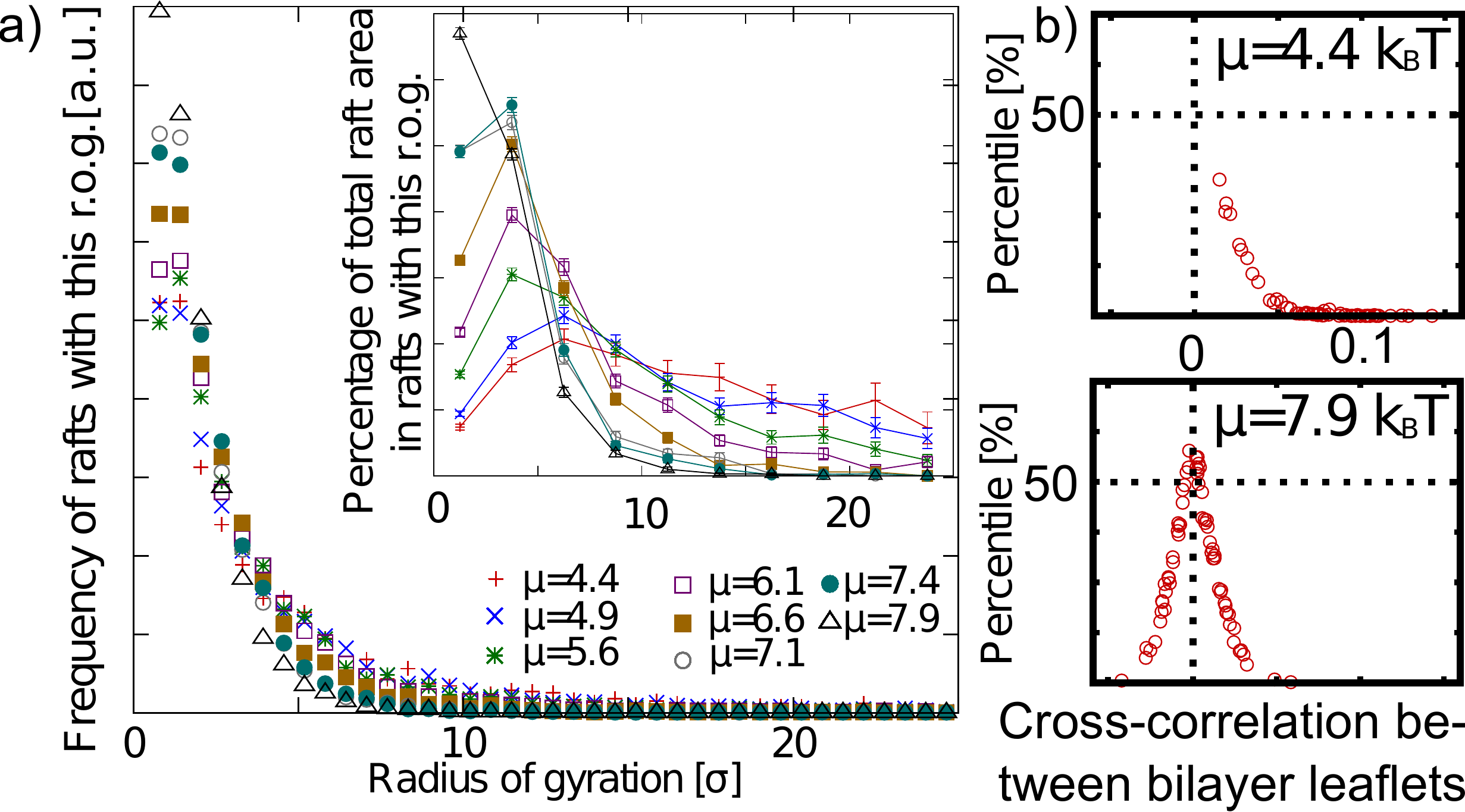}}
\caption{Characterization of raft domains. (a) Raft size distribution. The main 
panel shows the distribution of rafts with given radius of gyration. The inset 
shows the actual fraction of the raft area found in rafts of a given size. Lines 
are guides for the eye; {\footnotesize $\mu$} is given in units of 
{\footnotesize $k_BT$}. (b) Cross-correlation $K_i$ of configurations $i$ vs.\ 
percentile of conformations with randomly displaced monolayers, which have a 
higher correlation or anticorrelation $|\tilde{K}_i|$. Every point corresponds 
to an independent simulation configuration $i$. The more skewed a distribution 
is to the right side, the greater the mean (positive) correlation between rafts 
on both sides. The lower the percentile of a point, the less likely this 
particular value is coincidental. a.u., arbitrary units.} \label{fig:analysis_large}
\end{figure}

\begin{figure}
\centerline{\includegraphics[width=.3\textwidth]{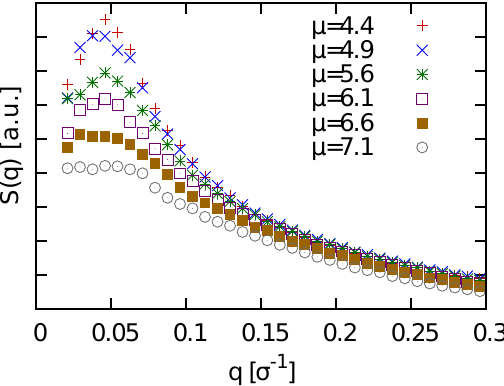}}
\caption{Radially averaged structure factor of the raft conformations for 
different values of $\mu$, with $\mu$ given in units of $k_BT$. a.u., arbitrary units.} 
\label{fig:structure_factor}
\end{figure}

\section{Elastic theory of raft stabilization by curvature} 
The main results of the simulations can be summarized as follows: (i) 
\lo~domains (rafts) of a finite nanoscale size are observed in two-component 
lipid bilayers, (ii) they are correlated across the membrane, i.e., rafts on the 
two leaflets tend to oppose each other, and (iii) the analysis of pressure 
profiles in small systems shows that \lo~monolayer domains have a propensity to 
bend inwards (a spontaneous curvature). However, large scale bending is 
prevented by the presence of the rafts on the opposing monolayer. These 
observations motivate the hypothesis that the elastic energy associated with the 
spontaneous curvature might be responsible for the finite size of the rafts (see 
Fig.~1).

To analyze this possibility, we consider a simple elastic model for two coupled 
monolayers with composition-dependent spontaneous curvature, which combines a 
model for mixed films by Leibler and Andelman \cite{LeiblerAndelman_87} with a 
bilayer model developed by Dan, Pincus, and coworkers \cite{DanPincus_93, DanBerman_94, 
Aranda-EspinozaBerman_96} to describe bilayer 
deformations near inclusions. For simplicity, and in contrast to the model proposed 
by Schick \cite{Schick_12}, we assume that the local compositions on opposing 
monolayers are strictly equal and do not induce bilayer bending. Bending and 
thickness deformations then decouple, and for planar membranes, the elastic free 
energy of monolayer thickness deformations can be written as follows
\cite{Aranda-EspinozaBerman_96, BranniganBrown_06, BranniganBrown_07, 
WestBrown_09}:
\begin{eqnarray}
\label{eq:free_energy}
F_{el} &=& \int {\rm d}^2 r \Big\{ 
\frac{k_c}{2} (\nabla^2 u)^2 
+ \frac{k_A}{2t_0^2} u^2 
\\ \nonumber &&
+ 2 k_c \big(\co + \frac{\zeta}{t_0} u\big) \nabla^2 u  
+ k_G \det(\partial_{ij} u) 
\Big\},
\end{eqnarray}
where $u({\bf r})$ denotes the local deviation from the mean monolayer thickness 
$t_0$, and the other parameters represent material properties of the membrane: 
the bilayer bending and compressibility modulus $k_c$ and $k_A$, the spontaneous 
monolayer curvature $\co$, an associated curvature-related parameter $\zeta$ 
\cite{Aranda-EspinozaBerman_96}, and the Gaussian rigidity of monolayers $k_G$. 
Eq.~(\ref{eq:free_energy}) holds for tensionless membranes as well as for 
membranes under tension~\cite{NederWest_10}. We have used it in the past to fit 
deformation profiles of one-component membranes in the vicinity of inclusions, 
with good results even on molecular length scales \cite{WestBrown_09, 
NederNielaba_12}.

We assume that the monolayer laterally phase separates into two phases, which 
are separated by narrow interfaces with a bare line tension $\lambda_0$. In 
principle, all membrane parameters ($t_0, k_c, k_A, \co, \zeta, k_G$) should 
depend on the local composition. For simplicity, however, we will assume that 
only the spontaneous curvature $\co$ makes a jump from one phase to the other. 
In that case, the final elastic energy after minimization can be written in the 
simple form (see Methods section)
\begin{equation}
\label{eq:energy_simple}
F_{el} = k_c \: \dc \: \int {\rm d}l \: {\bf n} \nabla u,
\end{equation}
where the line integral $\int {\rm d}l$ runs over all domain boundaries, ${\bf 
n}$ is the unit vector normal to the interface, and $\dc$ is the curvature 
mismatch, i.e., the difference of the spontaneous curvatures in the inner and 
the outer phase.

In many cases, this simple theory can be solved analytically. Details of the 
calculations are presented in the Methods section. For isolated plane 
interfaces, we obtain an elastic line energy (an elastic energy per boundary 
length $L$)
\begin{equation}
\label{eq:lambda_inf}
\lambda_{el}^{\infty} := {F_{el}}/{L} = - \xi k_c \: \dc^2 \: \big/ \sqrt{2(1-b)},
\end{equation}
which acts as an additive contribution to the total line tension, $\lambda_t =
\lambda_0 + \lambda_{el}^{\infty}$. Note that $\lambda_{el}^{\infty}$ is
negative. Here we have introduced the in-plane correlation length $\xi = ({k_c
t_0^2}/{k_A})^{1/4}$ and the dimensionless membrane parameter $b=2 \zeta
\xi^2/t_0$. Since $k_c$ should be roughly proportional to $k_A t_0^2$
\cite{RawiczOlbrich_00}, $\xi$ is of the order of the membrane thickness.
Inserting actual numbers for the fluid phase of one-component DPPC bilayers
from experiments, all-atom simulations, or simulations of our model
\cite{WestBrown_09}, one consistently obtains values around $\xi \sim
(0.9-1.4)$~nm. For the membrane parameter $b$, one obtains $b=0.65$ for our
DPPC model and $b=0.69$ for all-atom simulations of DPPC. Throughout this
paper, we assume $|b|<1$.

Since the elastic contribution $\lambda_{el}^{\infty}$ to the line tension is 
negative, the effect of elastic relaxation between two curvature-mismatched 
phases is similar to that of adding line-active surfactant agents. To assess its 
impact on the demixing transition, we analyze the scaling of $\lambda_0$ and 
$\lambda_{el}^{\infty}$ close to the critical demixing temperature $T_c$. The 
bare line tension vanishes according to $\lambda_0 \sim (T_c-T)^\nu$ with the 
critical exponent $\nu=1$, corresponding to the universality class of the two 
dimensional Ising model~\cite{HonerkampCicuta_08}. Likewise, the curvature 
mismatch $\dc$ will vanish upon approaching $T_c$, and it seems reasonable to 
assume that $\dc$ is proportional to the composition difference of the two 
phases, i.e., the order parameter of the demixing transition. The elastic ``line 
tension'' should therefore scale as $\lambda_{el}^{\infty} \propto \dc^2 \sim 
(T_c-T)^{2 \beta}$ with the 2D Ising exponent $\beta = 1/8$. Comparing the 
exponents for $\lambda_0$ and $\lambda_{el}^{\infty}$, we find that 
$\lambda_{el}^{\infty}$ dominates close to $T_c$, and the line tension becomes 
negative. Thus, macroscopic demixing is suppressed at $T_c$. The demixing 
transition is shifted to lower temperatures and gives way to a microemulsion or 
a modulated phase.

Next, we consider disks with finite diameter $D$ of one phase immersed in the 
other. In that case, the elastic line energy depends on $D$, and we obtain (see 
Methods section)
\begin{equation}
\label{eq:lambda_disk}
\lambda_{el}^{\mbox{\tiny disk}}(D) 
= \frac{F_{el}}{\pi D} = - \frac{\pi D k_c \dc^2}{2\sqrt{1-b^2}} \;
\Re\Big[  (\xi \alpha)^2 J_1(\alpha \frac{D}{2}) 
          H_1^{(1)}(\alpha \frac{D}{2}) \big],
\end{equation}
where $J_n$ and $H_n^{(1)}$ are Bessel and Hankel functions of the first kind. 
For comparison, we also consider the elastic line energy for isolated stripe 
domains of width $D$. It is given by
\begin{equation}
\label{eq:lambda_stripe}
\lambda_{el}^{\mbox{\tiny stripe}}(D) =
- \frac{\xi^2 \: k_c \: \dc^2}{\sqrt{1-b^2}}  \;
  \Re(\alpha (1-e^{i \alpha D}).
\end{equation}

\begin{figure}
\centerline{\includegraphics[width=.5\textwidth]{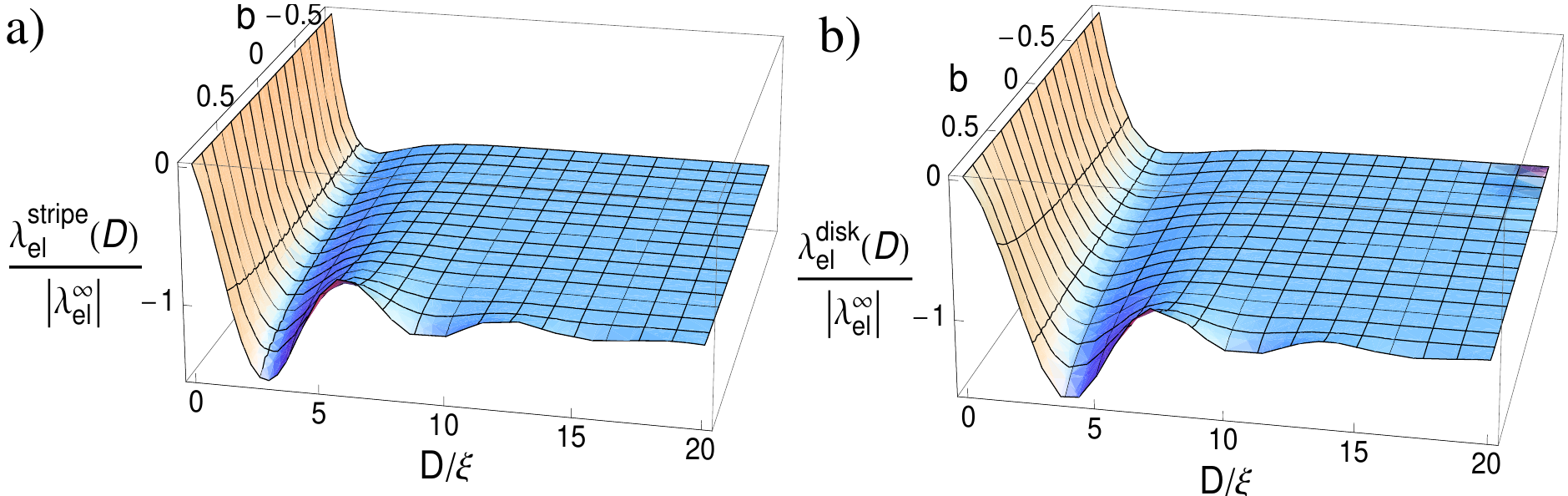}}
\caption{Rescaled elastic contribution to the line tension $\lambda_{el}(D)$, 
for (a) disk-shaped rafts of diameter $D$, and (b) stripe-shaped rafts of 
thickness $D$ vs.~$D$, in units of in-plane correlation length $\xi$ and 
membrane parameter $b$.} \label{fig:lambda}
\end{figure}

The functional dependence of $\lambda_{el}^{\mbox{\tiny disk}}(D)$ and 
$\lambda_{el}^{\mbox{\tiny stripe}}(D)$ on the domain size $D$ is illustrated in 
Fig.~6. As expected, both converge towards 
$\lambda_{el}^{\infty}$ for large $D$. At finite $D$, the behavior is 
nonmonotonic: Both line energies exhibit a minimum of similar depth (equal 
within 3 \%), which is weak for negative $b$ and becomes more pronounced as $b$ 
approaches $b \to 1$. The minimum is located at $D \sim 4 \xi$ for disks, and $D 
\sim 2-3 \xi$ for stripes. This result implies that domains of finite diameter 
become stable even before the asymptotic line tension vanishes: Thus the theory 
predicts a regime where the membrane is filled with small nanoscale domains, 
i.e., rafts.  Since disk-shaped and stripe-shaped domains have similar line 
energies, the actual shape depends on the composition of the membrane. At low 
\CC concentrations, disks will dominate; using $\xi \sim 1$~nm, the predicted 
characteristic raft size of around $4 \xi$ corresponds to a few nanometers. This 
is compatible with typical raft sizes observed in the simulation, see 
Fig.~4. Taking into account that $\sigma \sim 0.6$~nm, 
our estimate is at the lower end of the values typically suggested in the 
literature ($\sim 10-100$~nm).

We should add that the raft domains interact with each other, hence the theory 
actually predicts modulated phases with long-range order. Let us consider a 
system where the phase separation -- described by a demixing order parameter 
$\Phi$ -- is driven by a Ginzburg-Landau free energy functional of the form $ 
F_{\Phi} = \int {\rm d}^2 r \big\{ \frac{g}{2} (\nabla \Phi)^2 + f(\Phi) \big\} 
$ with $ f(\Phi) = \frac{r}{2} \Phi^2 - \frac{\gamma}{3!} \Phi^3 + 
\frac{\lambda}{4!} \Phi^4$ and let the spontaneous curvature $c_0$ depend 
linearly on $\Phi$ according to $c_0 = \Phi \hat{c}/\xi$. As shown in the 
Methods section, a homogeneous phase in this model becomes unstable with respect 
to modulations with wavelength $\xi$ at $g = 2 k_c \hat{c}^2/(1-b)$. Close to 
the spinodal, we recover the Landau-Brazovskii model, which provides a general 
framework for the description of phase transitions driven by a short-wavelength 
instability between a disordered phase and ordered phases \cite{Brazovskii_75}.

Since the same holds for the models for mixed membranes with bilayer curvature 
coupling mentioned in the introduction\cite{LeiblerAndelman_87, KumarGompper_99, 
Schick_12}, we expect the mean-field phase diagrams to be similar, with one 
important difference: In the bilayer coupling case, the characteristic 
wavelength $1/q^\star \sim \sqrt{k_c/\sigma}$ tends to be in the micrometer 
range and diverges for vanishing membrane tension $\sigma$, whereas here, the 
characteristic scale $\xi$ is in the nanometer range and independent of membrane 
tension. Consequently, the effect of fluctuations is expected to be much bigger 
in the present case. Fluctuations are known to shift the order-disorder 
transition and to stabilize a locally structured disordered phase {\em via} the 
Brazovskii mechanism \cite{Brazovskii_75}. The pattern formation then gives way 
to a microemulsion-type raft phase as observed in the simulations.
\section{Discussion}
To summarize, we have presented a coarse-grained simulation model for 
multicomponent lipid bilayer systems containing two types of lipids \PP and \CC 
with properties inspired by phospholipids and cholesterol. Our simulations show 
that this system forms thermodynamically stable nanoscale rafts of 
\mbox{\CC-enriched} \mbox{\lo-domains} surrounded by a sea of 
\mbox{\CC-depleted} \ld-phase. The in-plane structure factor features a peak at 
nonzero $q$, hence we have a microemulsion-like structure with a characteristic 
wavelength of around 12~nm. Furthermore, we have suggested a mechanism that 
stabilizes rafts of finite size. The mechanism is based on the idea that the 
spontaneous curvature of monolayer regions in the \lo- and \ld-phase should 
differ. We have established this for our model, and it seems likely that it is 
also the case in real membranes. The curvature mismatch then generates elastic 
interactions which suppress global phase separation and stabilize nanoscale 
structures.

If rafts are a disordered modulated structure, one might ask whether the
corresponding ordered modulated structure can be observed in nature as well.
Indeed, modulated ordered structures with a very similar length scale are found
in {\em one-component} membranes of phospholipids in the pre-transition region
between the fluid phase and the tilted gel phase ($L_{\beta'}$): The ripple
phase $P_{\beta'}$ \cite{KoynovaCaffrey_98}. It is generally observed in lipid
bilayers which exhibit a tilted gel phase. It is also observed in our model
\cite{LenzSchmid_07}, and the periodicity ($\sim 10$ nm) is comparable to the
characteristic length scale of our raft state. The structure of rippled
membranes is much more complicated than that of rafts. However, just like in
rafts, it involves alternating stretches of gel-like and liquid-like domains,
and simulations suggest that ripple formation is to a large extent driven by
lipid splay, i.e., by monolayer curvature \cite{LenzSchmid_07}. Thus we may
speculate that rafts and ripples represent just two sides of the same coin.
Curvature-mediated rafts might be a generic phenomenon in multicomponent
membranes, just like ripples are a generic phenomenon in one-component
membranes.

Unfortunately, rafts are much more difficult to study experimentally than
ripples due to the lack of long-range order, and due to their subsecond
lifetimes~\cite{JacobsonMouritsen_07}. The structure factor describing the
distribution of domains on the nanoscale could possibly be measured by X-ray
diffraction experiments on aligned multi-lamellar membranes (``membrane
stacks'') of the relevant composition, either in reflection or transmission
\cite{Salditt_05, AeffnerReusch_12}. This would allow one to test whether the
in-plane structure factor has a peak at nonzero wave vector in type~I
mixtures as predicted by our model [Fig.~5]. In
addition, new super resolution microscopy techniques
\cite{EggelingRingemann_09} might provide ways to visualize rafts in free
membranes on the scale of a few nanometers.

The curvature mechanism proposed here of course does not exclude other 
mechanisms of raft formation such as those discussed in the introduction. Many 
mechanisms might compete in nature. In particular, it should be interesting to 
study the interplay of curvature-mediated rafts and lipid-protein interactions 
\cite{NederNielaba_12, YethirajWeisshaar_07} in future work.

\begin{footnotesize}
\subsection*{Methods}
\subsubsection*{Coarse-grained simulation model} 
\label{methods_model}
The model is defined in terms of the length unit {\footnotesize $\sigma \approx
0.6 \text{nm}$} and the energy unit {\footnotesize $\varepsilon \approx
0.36\cdot 10^{-20} \text{J}$} \cite{WestSchmid_10}.  ``Phospholipids'' 
\mbox{(\MPP)}
are represented by simple flexible chains of beads with a hydrophilic head and
a hydrophobic tail, which self-assemble in the presence of structureless
solvent beads \cite{LenzSchmid_05}. ``Cholesterol'' molecules (\MCC)
have the same basic structure, but they are shorter and stiffer except for one
flexible end. All lipids are linear chains of six tail beads attached to one
head bead, connected by finite extension nonlinear elastic (FENE) springs with spring constant
{\footnotesize $k_b = 100 \frac{\varepsilon}{\sigma^2}$}, equilibrium bond
lengths {\footnotesize $r_0=0.7 \sigma$} (\MPP lipid) and {\footnotesize $r=0.6
\sigma$} (\MCC lipid), and logarithmic cutoffs at {\footnotesize $\Delta
r_{\text{max}}=0.2\sigma$} (\MPP) and {\footnotesize $\Delta
r_{\text{max}}=0.15\sigma$} (\MCC). Consecutive bonds in a chain with angle
{\footnotesize $\Theta$} are subject to a stiffness potential {\footnotesize
$U_{\text{BA}}(\Theta)=k_\theta (1-\cos(\Theta))$} with stiffness constant
{\footnotesize $k_\theta=4.7 \epsilon$} (\MPP lipids), {\footnotesize
$k_\theta=100 \epsilon$} (\MCC lipids, first four angles), and {\footnotesize
$k_\theta=4.7 \epsilon$} (\MCC lipid, last angle). Beads that are not directly
bonded with each other interact via a Lennard-Jones potential
{\footnotesize $U_{\text{LJ}}(r/\varsigma) =
\epsilon_{\text{LJ}}\big((\frac{\varsigma}{r})^{12}-
2(\frac{\varsigma}{r}t)^6\big)$}, which is truncated at a cutoff radius {\footnotesize $r_c$}
and shifted such that it remains continuous. At {\footnotesize $r_c =  1$}, one recovers the
purely repulsive Weeks-Chandler-Anderson potential \cite{WCA}. The interaction parameters for pairs
of \MPP or \MCC beads (head or tail) and solvent beads are given by

\smallskip
\newcommand{\EMBEDDED}{ON}
\begin{center}
\ifdefined\EMBEDDED
\else
\documentclass{pnastwo}
\usepackage{PNAStwoF}

\newcommand{\MPP}{{\em \footnotesize P }}
\newcommand{\MCC}{{\em \footnotesize C }}

\begin{document}
\fi

\begin{small}
\begin{tabular}{cccc}
{\footnotesize bead type-bead type}&$\epsilon/\varepsilon$ &$\varsigma/\sigma$&$r_c/\varsigma$\\
\hline
{\footnotesize head(any)-head(any)} & 1.0 & 1.1 & 1.0 \\
{\footnotesize head(any)-tail(any)} &1.0 &1.05 &1.0\\
{\footnotesize head(any)-solvent}  & 1.0 & 1.1 & 1.0\\
{\footnotesize tail(\MPP)-tail(\MPP)} & 1.0 & 1.0 & 2.0 \\
{\footnotesize tail(\MPP)-tail(\MCC)} & 1.0 & 1.0 & 2.0 \\
{\footnotesize tail(\MCC)-tail(\MCC)} & 0.9 & 1.0 & 2.0 \\
{\footnotesize tail(any)-solvent} & 1.0 & 1.05 &1.0\\
{\footnotesize solvent-solvent} & 0 & & 
\end{tabular}
\end{small}

\ifdefined\EMBEDDED
\else
\end{document}
\fi

\end{center}
\smallskip

Hence all non-bonded interactions except the tail-tail interactions are repulsive,
and the attraction between \MCC tail-beads is weaker than that between other tail beads.

The model was studied by Monte Carlo simulations at constant pressure
{\footnotesize $P=2 \varepsilon/\sigma^3$} and constant zero surface tension in
a fluctuating box of variable size and shape \cite{SchmidDuechs_07}. The total
number of lipids was kept fixed. The composition was sometimes allowed to
fluctuate (semi-grandcanonical ensemble). In that case, semi-grandcanonical
moves were implemented by means of configurational bias Monte Carlo moves
\cite{FrenkelSmit_book}, during which the identity of a lipid was switched
between \MPP and \MCC. In the canonical simulations, the same moves can be used
as virtual moves to determine the chemical potential difference {\footnotesize
$\mu$} for \MPP and \MCC chains.

\subsubsection*{Data analysis}
\label{methods_analysis}
In order to analyze configurations such as that shown in Fig.~3~(a), we map each monolayer onto a discrete
grid {\footnotesize $\chi_{xy}$}, where {\footnotesize $\chi=1$} stands for 
``raft'' and 
{\footnotesize $\chi=0$} for ``non-raft''. This is done with the following 
algorithm:
\begin{enumerate}
\item Assign chains to upper and lower bilayer leaflet according to their head-tail orientation.
\item For each layer sort chains into quadratic bins of side length {\footnotesize $1.5\sigma$} 
according to the {\footnotesize{$xy$}}-Position of the head-bead.
\item Calculate the number density {\footnotesize $\rho$}, the \MCC density 
{\footnotesize $\rho_c$} and the nematic order {\footnotesize $S$} for each bin.
\item At each vertex of the lattice take the mean value of each observable in the surrounding bins.
\item If {\footnotesize $\rho\cdot\rho_c\cdot S>0.15\sigma^{-4}$} the membrane at the vertex 
position is considered to be in the \lo-Phase.
\item Apply the Density Based Spatial Clustering of Applications with Noise
(DBSCAN) cluster detection algorithm\cite{Ester_96} with Eps=3{\footnotesize$\sigma$} 
and minimum number of points (MinPts) = 3.  The clusters are identified as lipid rafts.
\end{enumerate}
The normalized cross-correlations {\footnotesize $K_i$} of opposing monolayers of a configuration
{\footnotesize $i$} are calculated according to
{\footnotesize $K_i = \frac 1N\sum_{x,y}{ (\chi_{xy}^{it} - \bar{\chi}^{it}) (\chi_{xy}^{ib} - \bar{\chi}^{ib})} \big/
     {\sigma_{it}\sigma_{ib}}$},
where {\footnotesize $\bar{\chi}^{it}$} and {\footnotesize $\bar{\chi}^{ib}$} are the mean values of
{\footnotesize $\chi_{xy}^{it}$} and {\footnotesize $\chi_{xy}^{ib}$}, respectively, 
{\footnotesize $\sigma_{it}$} and {\footnotesize $\sigma_{ib}$} are their standard deviations, and 
{\footnotesize $N$} is the number of lattice
points. To analyze whether a correlation {\footnotesize $K_i$} is significant, 
we take the same configuration {\footnotesize $i$}, displace one of the leaflets by a random offset
with respect to the other ({\footnotesize $\tilde{\chi}^{ib}_{xy} = \chi^{ib}_{x+r_x,y+r_y}$}),
and compare the new correlation {\footnotesize $\tilde{K}_i$} with {\footnotesize $K_i$}. 

\subsubsection*{Solution of the elastic theory}
The elastic free energy {\footnotesize $F_{el}$} (Eq.\ (\ref{eq:free_energy}))
is minimized with respect to {\footnotesize $u({\bf r})$} in the bulk (inside
phase separated domains) and to the boundary values of {\footnotesize $u$} and
the normal derivative {\footnotesize ${\bf n} \nabla u$} at the domain
boundaries. We note that the latter must be continuous across the boundaries,
otherwise {\footnotesize $F_{el}$} diverges. This results in the Euler-Lagrange
equation {\footnotesize $\xi^4 \: \nabla^4 u + 2 b \: \xi^2 \: \nabla^2 u + u =
0$} in the bulk, and
in boundary conditions at the interfaces between domains: {\footnotesize ${\bf
n}\nabla (\nabla^2 u)$} must be continuous, and {\footnotesize $\nabla^2 u$}
jumps by {\footnotesize $2 \dc$} at the boundaries. Inserting the
Euler-Lagrange equations and the boundary condition in Eq.\
(\ref{eq:free_energy}) gives the simplified expression
(\ref{eq:energy_simple}).

For stripe domains of width {\footnotesize $D$} with boundaries at {\footnotesize $z = \pm
D/2$}, the deformation profile {\footnotesize $u(z)$} satisfying the Euler-Lagrange
equation and the boundary conditions is given by
\begin{scriptsize}
\begin{equation}
\label{eq:u_stripe}
u(z) = - \frac{2 \xi^2 \:\dc}{\sqrt{1-b^2}}  \left\{
\begin{array}{ll}
\Re\big[ i e^{i \alpha D/2} \cos(\alpha z)\big] 
&:\;  0 < z < D/2 \\
\Re\big[ \sin(-\alpha D/2) \: e^{i \alpha z} \big] 
&:\;  z > D/2 
\end{array}
\right.
\end{equation}
\end{scriptsize}
with {\scriptsize $\alpha = \sqrt{b+i\sqrt{1-b^2}}/\xi$}, and {\footnotesize $u(-z) = u(z)$}.
Two isolated plane boundaries are obtained in the limit {\footnotesize $D \to \infty$}.
Finally, the radially symmetric analytical solution for disk domains is 
\begin{scriptsize}
\begin{equation}
\label{eq:u_disk}
u(r) = \frac{ \dc }{\sqrt{1-b^2}}  \: 
\frac{\pi D \xi}{2}
\left\{ \begin{array}{cl}
\Re \big[\xi \alpha \: J_0(\alpha r) \: H_1^{(1)}(\alpha D/2)\big] \; &: \: r < R \\
\Re \big[\xi \alpha \: J_1(\alpha D/2) \: H_0^{(1)}(\alpha r)\big] \; &: \: r > R 
\end{array} \right.,
\end{equation}
\end{scriptsize}
The profile (\ref{eq:u_disk}) satisfies the 
boundary conditions by virtue of the identity
{\scriptsize $J_0(z) H_1^{(1)}(z) - J_1(z) H_0^{(1)}(z) = - 2 i/\pi z$.}
Inserting the expressions (\ref{eq:u_stripe}) and
(\ref{eq:u_disk}) into Eq.\ (\ref{eq:energy_simple}) gives
(\ref{eq:lambda_stripe}), (\ref{eq:lambda_disk}), and
(\ref{eq:lambda_inf}) (in the limit {\footnotesize $D \to \infty$}).

\subsubsection*{Derivation of the Landau-Brazovskii model} 
To analyze the situation
close to the spinodal, we minimize {\footnotesize $F = F_{el}+F_{\Phi}$} 
for {\footnotesize $c_0 =
\phi \hat{c}/\xi$} with respect to {\footnotesize $u$}.  In wavevector space
{\footnotesize ${\mathbf{q}}$}, this gives {\footnotesize $u_{\mathbf{q}} =
\Phi_{\mathbf{q}} \cdot 2 \hat{c} \xi \: \chi\big((q \xi)^2\big) $} with
{\footnotesize $\chi(x) = x/(x^2-2 b x + 1)$}. Inserting this in
{\footnotesize $F$} and omitting
boundary terms, we obtain {\footnotesize $F = \frac{1}{2} \sum_{\mathbf{q}}
|\Phi_{\mathbf{q}}|^2 q^2 g_{\mbox{\tiny eff}}(q^2) + \int {\rm d}^2 r \:
f(\Phi)$} with {\footnotesize $g_{\mbox{\tiny eff}}(q^2) = g - 4 k_c \hat{c}^2
\chi\big((q \xi)^2\big)$}.  The function {\footnotesize $g_{\mbox{\tiny eff}}(q^2)$}
has a minimum at {\footnotesize $q^* = 1/\xi$}, hence a homogeneous phase (with
{\footnotesize $\Phi = $}constant) becomes unstable at {\footnotesize
$g_{\mbox{\tiny eff}}(q^*) = 0$}, i.e., {\footnotesize $g = g^* := 2 k_c
\hat{c}^2/(1-b)$}. Close to the spinodal, contributions {\footnotesize $q \sim
q^*$} dominate, hence we expand {\footnotesize $q^2 g_{\mbox{\tiny eff}}(q^2)$}
about {\footnotesize ${q^*}^2$} up to second order.  This finally gives a free
energy expression of the Landau-Brazovskii form
{\footnotesize $
F = \int {\rm d}^2 r \big\{ 
\frac{\Gamma}{2} (\Delta + q_0^2)^2 \Phi^2 
+ \frac{\tau}{2} \Phi^2
- \frac{\gamma}{3!} \Phi^3
+ \frac{\lambda}{4!} \Phi^4
\big\}
$}
with 
{\footnotesize $q_0^2 = {q^*}^2 \big(1+(1-b)(1-\frac{g}{g^*})\big)$}, 
{\footnotesize $\tau = r + \frac{1}{2} (g-g^*)(q_0^2+{q^*}^2)$}, and 
{\footnotesize $\Gamma = g^* / [2 (1-b) {q^*}^2]$}.
\end{footnotesize}

\bigskip

\section*{Acknowledgments}
We thank S. Keller, T. Salditt, and M. Schick for valuable discussions. This 
work was supported by the German Science Foundation within the collaborative research center SFB-625. 
Simulations were carried out at the John von Neumann Institute for Computing (NIC) J\"ulich, and 
the Mogon Cluster at Mainz University.


\begin{thebibliography}{10}

\bibitem{SimonsvanMeer_88}
Simons~K, van~Meer~G (1988)
{\em Lipid sorting in epithelial cells}.
Biochemistry 27:6197--6202.

\bibitem{SimonsIkonen_97}
Simons~K, Ikonen~E (1997)
{\em Functional rafts in cell membranes}.
Nature  387:569--572.

\bibitem{ZurzolovanMeer_03} Zurzolo~C, van~Meer~G, Mayor~S (2003)
{\em The order of rafts}.
EMBO reports 4:1117--1121.

\bibitem{Pike_06} Pike~LJ (2006)
{\em Rafts defined: A report on the keystone symposium on lipid rafts and cell function}.
J Lipid Res 47:1597--1598.

\bibitem{Hancock_09} Hancock~JF (2006)
{\em Lipid rafts: contentious only from simplistic standpoints}.
Nat Rev Mol Cell Biol 7:456--462.

\bibitem{Leslie_11}
Leslie~M (2011)
{\em Do lipid rafts exist?}.
Science 334:1046--1047.

\bibitem{VerebSzollosi_03} Vereb~G et al. (2003)
{\em Dynamic, yet structured: The cell membrane three decades after the Singer-Nicolson model}.
PNAS 100: 8053--8058.

\bibitem{VeatchKeller_05} Veatch~SL, Keller~SL (2005)
{\em Seeing spots: Complex phase behavior in simple membranes}.
Biochim Biophys Acta 1746:172--185.

\bibitem{BrownLondon_98} Brown~DA, London~E (1998) 
{\em Functions of lipid rafts in biological membranes}.
Annu Rev Cell Dev Biol 14:111--136.

\bibitem{JacobsonDietrich_99} Jacobson~K, Dietrich~C (1999)
{\em Looking at lipid rafts?}.
Trends in Cell Biology 9:87--91. 

\bibitem{Edidin_03} Edidin~M (2003)
{\em The state of lipid rafts: From model membranes to cells}.
Annu Rev Biophys Biomol Struct 32:257--283.

\bibitem{Pike_03} Pike~LJ (2003)
{\em Lipid rafts: Bringing order to chaos}.
J Lipid Res 44:655--667.

\bibitem{SimonsVaz_04} Simons~K, Vaz~WLC (2004)
{\em Model systems, lipid rafts, and cell membranes}.
Annu Rev Biophys Biomol Struct 33:269--295.

\bibitem{ZeydaStulnig_06} Zeyda~M, Stulnig~TM (2006)
{\em Lipid rafts \& Co.: An integrated model of membrane organization in T cell activation}.
Progr in Lipid Research 45:187--202.

\bibitem{Hanzal-BayerHancock_07} Hanzal-Bayer~MF, Hancock~JF (2007)
{\em Lipid rafts and membrane traffic},.
FEBS Letters 581:2098--2104.

\bibitem{LingwoodSimons_10}
Lingwood~D, Simons~K (2010)
{\em Lipid rafts as a membrane-organizing principle}.
Science 327:46--40.

\bibitem{PralleKeller_00}
Pralle~A, Keller~P, Florin~EL, Simons~K, H\"orber~JKH (2000)
{\em Sphingolipid-cholesterol rafts diffuse as small entities in the plasma membrane of mammalian cells}.
J Cell Biol 148:997--1007. 

\bibitem{ZachariasViolin_02}
Zacharias~DA, Violin~JD, Newton~AC, ~Tsien~RY (2002)
{\em Partitioning of lipid-modified monomeric GFPs into membrane microdomains of live cells}.
Science 296:913--916.

\bibitem{BrownRose_92}
Brown~DA, Rose~JK (1992)
{\em Sorting of GPI-anchored proteins to glycolipid-enriched
membrane subdomains during transport to the apical cell surface}.
Cell 68:533--544.

\bibitem{TurnerSens_05} Turner~MS, Sens~P, Socci~ND (2005)
{\em Nonequilibrium raftlike membrane domains under continuous recycling},
Phys Rev Lett 95(16):168301

\bibitem{YethirajWeisshaar_07} Yethiraj~A, Weisshaar~JC (2007)
{\em Why are lipid rafts not observed in vivo?}.
Biophys J 93:3113--3119.

\bibitem{VeatchKeller_03} Veatch~SL, Keller~SL (2003)
{\em Separation of liquid phases in giant vesicles of ternary mixtures of
phospholipids and cholesterol}.
Biophys J 85:3074--3083.

\bibitem{VeatchKeller_05b} Veatch~SL, Keller~SL (2005)
{\em Miscibility phase diagrams of giant vesicles containing sphingomyelin}.
Phys Rev Lett 94:148101-1--4

\bibitem{VeatchCicuta_08} Veatch~SL et al. (2008) 
{\em Critical fluctuations in plasma membrane vesicles}.
ACS Chemical Biology 3:287--293.

\bibitem{VeatchSoubias_07} Veatch~SL, Soubias~O, Keller~SL, Gawrisch~K (2007)
{\em Critical fluctuations in domain-forming lipid mixtures}.
PNAS 104:17650--17655.

\bibitem{Honerkamp-SmithVeatch_09}
Honerkamp-Smith~AR, Veatch~SL, Keller~SL (2009)
{\em An introduction to critical points for biophysicists: Observations of compositional
heterogeneity in lipid membranes}.
Biochim Biophys Acta 1788:53--63.

\bibitem{GompperSchick_book} Gompper~G, Schick~M (1994) in
{\em Phase transitions and critical phenomena: 
Self-Assembling Amphiphilic Systems}, (Academic Press, London, UK), Vol 16, pp 16--18

\bibitem{Schick_12} Schick~M (2012)
{\em Membrane heterogeneity: Manifestation of a curvature-induced microemulsion}. 
Phys Rev E 85:031902-1--4.

\bibitem{SafranPincus_90} Safran~SA, Pincus~P, Andelman~D (1990)
{\em Theory of spontaneous vesicle formation in surfactant mixtures}.
Science 248:354--356.

\bibitem{LeiblerAndelman_87} Leibler~S, Andelman~D (1987)
{\em Ordered and curved meso-structures in membranes and amphiphilic films}.
J Physique 48:2013--2018.

\bibitem{HardenMacKintosh_94} Harden~JL, MacKintosh~FC (1994)
{\em Shape transformations of domains in mixed-fluid films and
bilayer membranes}.
Europhys Lett 28:495--500.
 
\bibitem{KumarGompper_99} Kumar~PBS, Gompper~G, Lipowsky~R (1999)
{\em Modulated phases in multicomponent fluid membranes}.
Phys Rev E 60:4610--4618.

\bibitem{BaumgartHess_03} Baumgart~T, Hess~ST, Webb~WW (2003)
{\em Imaging coexisting fluid domains in biomembrane models
coupling curvature and line tension}.
Nature 425:821--824.

\bibitem{BrewsterPincus_09} Brewster~R, Pincus~PA, Safran~SA (2009)
{\em Hybrid lipids as a biological surface-active component}.
Biophys J 97:1087 -- 1094.

\bibitem{YamamotoBrewster_10} Yamamoto~T, Brewseter~R, Safran~SA (2010)
{\em Chain ordering of hybrid lipids can stabilize domains in
saturated/hybrid/cholesterol lipid membranes}.
EPL 91:28002-1--6.

\bibitem{IpsenKarlstrom_87} Ipsen~JH, Karlstr\"om~G, Mouritsen~OG,
Wennerstr\"om~H, Zuckermann~MJ (1987)
{\em Phase equilibria in the phosphatidylcholine-cholesterol system}.
Biochim Biophys Acta 905:162--172.

\bibitem{SankaramThompson_91} Sankaram~MB, Thompson~TE (1991)
{\em Cholesterol-induced fluid-phase immiscibility in membranes}.
PNAS 88:8686--8690.

\bibitem{AlmeidaVaz_92} Almeida~PFF, Vaz~WLC, Thompson~TE (1992)
{\em Lateral diffusion in the liquid phases of dimyristoylphophatidylcholine/cholesterol lipid bilayers: A free volume analysis}.
Biochemistry 31:6739--6747.

\bibitem{LindblomOradd_09} Lindblom~G, Or\"add~G (2009)
{\em Lipid lateral diffusion and membrane heterogeneity}.
Biochim Biophys Acta:234--244.

\bibitem{Feigenson_09} Feigenson~GW (2009)
{\em Phase diagrams in lipid domains in multicomponent lipid bilayer mixtures}.
Biochim Biophys Acta:47--52.

\bibitem{CollinsKeller_08} Collins~MD, Keller~SL (2008)
{\em Tuning lipid mixtures to induce or suppress domain formation
across leaflets of unsupported asymmetric bilayers}.
PNAS 105:124--128.

\bibitem{SchmidDuechs_07} Schmid~F, D\"uchs~D, Lenz~O, West~B (2007)
{\em A generic model for lipid monolayers, bilayers, and membranes}.
Comp Phys Comm 1177:168--171.

\bibitem{LenzSchmid_07} Lenz~O, Schmid~F (2007).
{\em Structure of symmetric and asymmetric ripple phases in lipid bilayers}.
Phys Rev Lett 98:058104-1--4.

\bibitem{WestSchmid_10} West~B, Schmid~F (2010)
{\em Fluctuations and elastic properties of lipid membranes in the fluid
and gel state: A coarse-grained Monte Carlo study}.
Soft Matter 6:1275--1280.

\bibitem{WestBrown_09} West~B, Brown~FLH, Schmid~F (2009)
{\em Membrane-protein interactions in a generic coarse-grained model for lipid bilayers}.
Biophys J 96:101--115.

%
%
\bibitem{Waheed_12} Waheed~Q (2012) {\em Molecular Dynamics Simulations of Biological Membranes
}. Doctoral Thesis (Royal Institute of Technology, Stockholm, Sweden).


\bibitem{WaheedTjornhammar_12} Waheed~Q, Tj\"ornhammar~R, Edholm~O (2012)
{\em Phase transitions in coarse grained lipid bilayers containing cholesterol
by molecular dynamics simulations}.
Biophys J 103:2125-2133.

\bibitem{Safran_book} Safran~SA (1994) in
{\em Statistical Thermodynamics of Surfaces, Interfaces, and Membranes}, (Addison-Wesley, Reading), pp 193--200.

\bibitem{DanPincus_93} Dan~N, Pincus~P, Safran~SA (1993)
{\em Membrane induced interactions between inclusions}.
Langmuir 9:2768--2771.

\bibitem{DanBerman_94} Dan~N, Berman~A, Pincus~P, Safran~SA (1994)
{\em Membrane induced interactions between inclusions}.
J de Physique II 4:1713--1725.

\bibitem{Aranda-EspinozaBerman_96} H.\ Aranda-Espinoza, A.\ Berman, N.\ Dan,
P.\ Pincus, S.\ Safran (1996)
{\em Interaction between inclusions embedded in membranes}.
Biophys J 71:648--656.

\bibitem{BranniganBrown_06} Brannigan~G, Brown~FLH (2006)
{\em A consistent model for thermal fluctuations and protein-induced
deformations in lipid bilayers}.
Biophys J 90:1501--1520.

\bibitem{BranniganBrown_07} Brannigan~G, Brown~FLH (2007)
{\em Contributions of Gaussian curvature and nonconstant lipid volume
to protein deformations of lipid bilayers}.
Biophys J 92:864--867.

\bibitem{NederWest_10} Neder~J, West~B, Nielaba~P, Schmid~F (2010)
{\em Coarse grained simulations of membranes under tension}.
J Chem Phys 132:115101-1--12.

\bibitem{NederNielaba_12} Neder~J, Nielaba~P, West~B, Schmid~F (2012)
{\em Interactions of membranse with coarse-grain proteins: A comparison}.
New J Physics 14:125017-1--24.

\bibitem{RawiczOlbrich_00} Rawicz~W, Olbrich~KC, McIntosh~T, Needham~D,
Evans~E (2000) 
{\em Effect of chain length and unsaturation on elasticity of lipid bilayers}.
Biophys J 79:328--339.

\bibitem{HonerkampCicuta_08} Honerkamp-Cicuta~AR et al. (2008)
{\em Line tensions, correlation lengths, and critical exponents in lipid 
membranes near critical points}.
Biophys J 95:236--246.

\bibitem{Brazovskii_75} Brazovskii~SA (1975)
{\em Phase transitions of an isotropic system to a nonuniform state}.
Soviet Physics JETP 41(1):85--89.

\bibitem{KoynovaCaffrey_98} Coynova~R, Caffrey~M (1998)
{\em Phases and phase transitions of the phosphatidylcholines}.
Biochim Biophys Acta 1376:91--145.

\bibitem{JacobsonMouritsen_07} Jacobson~K, Mouritsen~O, Anderson~R (2007)
{\em Lipid rafts: at a crossroad between cell biology and physics}.
Nature Cell Biology 9:7--14.

\bibitem{Salditt_05} Salditt~T (2005)
 {\em Thermal fluctuations and stability of solid-supported lipid membranes}.
J Phys: Cond Matter 17:R287--R314.

\bibitem{AeffnerReusch_12} Aeffner~S, Reusch~T, Weinhausen~B, Salditt~T (2012)
{\em Energetics of stalk intermediates in membrane fusion are controlled by lipid composition}.
PNAS 109,25:E1609--E1618

\bibitem{EggelingRingemann_09} Eggeling~C et al. (2009)
{\em Direct observation of the nanoscale dynamics of membrane lipids in a living cell}.
Nature 457:1159--1162.

\bibitem{LenzSchmid_05} Lenz~O, Schmid~F (2005)
{\em A simple computer model for liquid lipid bilayers}.
J Mol Liquids 117:147--152.

\bibitem{WCA} Weeks~JD, Chandler~J, Andersen~HC (1971)
{\em Role of repulsive forces in forming the equilibrium structure of simple liquids}.
J Chem Phys 54:5237--5247.

\bibitem{FrenkelSmit_book} Frenkel~D, Smit~B (2001) in 
{\em Understanding molecular simulation: From algorithms to applications}. (Academic), 2nd Ed, pp 336--350

\bibitem{Ester_96} Ester~M, Kriegel~H, Sander~J, Xu~X (1996)
{\em A density-based algorithm for discovering clusters in large spatial databases with noise}.
 Proceedings of the Second International Conference on Knowledge Discovery 
 and Data Mining (AAAI Press, Menlo Park, CA), 226-231.


\end{thebibliography}
\end{document}